\documentclass[a4paper,12pt,reqno]{amsart}

\usepackage[centertags]{amsmath}
\usepackage{amsfonts}
\usepackage{amssymb}
\usepackage{amsthm}
\usepackage{newlfont}
\usepackage{stmaryrd}
\usepackage{mathrsfs}
\usepackage{euscript}
\usepackage{bbm}
\usepackage{graphicx}

\usepackage{color}

\theoremstyle{plain}
  \newtheorem{theorem}{Theorem}
  
  \newtheorem{proposition}[theorem]{Proposition}
  
\theoremstyle{definition}

\theoremstyle{remark}

\newcommand{\bbR}{{\mathbb R}}

\newcommand{\opunit}{\text{1}\kern-0.22em\text{l}}

\newcommand{\id}{\textrm{d}}

\begin{document}

{{\Large {\bf On a response formula and its\\

interpretation}}}\\

{\large Christian Maes
and Bram Wynants}\\

{\tt Instituut voor Theoretische Fysica, K.U.Leuven, Belgium}

\vspace{4,5 cm} \noindent {\bf Abstract}:  We present a
 physically inspired generalization of equilibrium response
formul{\ae}, the fluctuation-dissipation theorem, to Markov jump
processes possibly describing interacting particle systems
out-of-equilibrium, following the recent work of \cite{prl,bmw}.
Here, the time-dependent perturbation adding a potential $V$ with
small amplitude $h_t$  changes the rates $W(x,y)$ for the
transition $x\rightarrow y$ into
\[
W_t(x,y) = W(x,y)\,e^{h_t\,(bV(y)-aV(x))}
\]
as first considered by Diezemann, \cite{diez}; $a,b$ are
constants. We observe that the linear response relation shows a
reciprocity symmetry in the nonequilibrium stationary regime and
we interpret the connection with dynamical fluctuation theory.

\vspace{3,5cm} \noindent{\it Dedicated to the 15th birthday of\\
{\bf Markov Processes and Related Fields}.}

\newpage

\section{Hurrah}
The present paper is devoted to an important theme of statistical
thermodynamics and system theory.  It deals with the response of a
system to an external stimulus.  More specifically, we are
interested in the linear response to an energy impulse applied to
nonequilibrium systems and generalizing the celebrated
fluctuation-dissipation theorem. That question comes up in the
general construction of nonequilibrium statistical mechanics, but
has possible applications in a variety of contexts. It is not
clear yet whether the answer that we give here and that was
presented in a more restricted sense in the physics literature
\cite{prl,bmw}, is operationally useful. We do however attempt
some interpretation directed towards dynamical fluctuation theory.
The mathematical origin of all this is the theory of stochastic
processes, here in its most simple representation for Markov jump
processes on finite alphabets. The very fact that physically
interesting relations can possibly be suggested already from the
elementary mathematical theory of Markov processes is rather
encouraging for the Markov field. That is then our contribution to
the celebration of a young journal devoted to that subject, hurrah
and many years to come.

\section{Response in equilibrium}
Relations between fluctuations, response
behavior and dissipation in equilibrium systems have been
obtained and applied throughout the development of statistical
mechanics in the 20th century, \cite{ku}.  Quite often textbooks treat linear response in a
 quantum mechanical context when applied to discrete systems such
  as spins or particles hopping on a lattice.
 The method for that equilibrium fluctuation-dissipation theorem
 is first order perturbation theory on
time-dependent Liouville--von Neumann equations.\\
  Here we
consider stochastic evolutions, Markov jump processes; we give a
more probabilistic treatment and
corresponding statistical mechanical interpretation.  The mathematics remains elementary.\\

\subsection{Equilibrium dynamics}
  Let us consider a
simple situation, which falls in the context of the present
discussion.  Take an Ising spin system on a finite graph
$(\Lambda,\sim)$; at each vertex $i\in \Lambda$ there is a spin
$\sigma(i)=\pm 1$. A spin flip Markov dynamics on the
configurations $\sigma \in \{+1,-1\}^\Lambda\equiv K$ has
possible transitions $\sigma\rightarrow \sigma^j$ where
$\sigma^j(i)=\sigma(i)$ for $j\neq i$ and $ \sigma^j(j)=-\sigma(j)$ is
the new  configuration with the spin flipped at vertex $j$.  Physically, we
imagine that there is a thermal reservoir perhaps in the form of
lattice vibrations or of electronic degrees of freedom attached to
the system so that for each transition $\sigma\rightarrow
\sigma^j$ there is an energy exchange $U(\sigma^j) - U(\sigma)$
with and an entropy flux $[U(\sigma) - U(\sigma^j)]/T$ in the
reservoir at equilibrium temperature $T$. There is no need here to
specify that energy function $U(\sigma)$. As long as we assume the system is only
in contact with the outside world through this one temperature bath,
there should be a reversible
stationary distribution $\rho$ giving probabilities
\begin{equation}\label{eq}
\rho(\sigma) = \frac 1{Z} e^{-\beta U(\sigma)}
\end{equation}
to the  spin configurations. We call it the equilibrium
distribution. The reversibility is expressed by the condition of
detailed balance
\[
W(\sigma,\sigma^j)\,\rho(\sigma)  =
W(\sigma^j,\sigma)\,\rho(\sigma^j),\quad \mbox{for all } j\in
\Lambda, \sigma \in K
\]
for the rates $W(\sigma,\sigma^j)$ of the transitions $\sigma \to \sigma^j$.
Remark that all static
properties of the system depend only on that part of the rates that is determined
by the detailed balance condition. In other words, as long as the rates satisfy
\[ \frac{W(\sigma,\sigma^j)}{W(\sigma^j,\sigma)} = e^{-\beta[U(\sigma^j)-U(\sigma)]} \]
they will simulate the correct physical properties of the system.
Explicitly, the transition rates of the stochastic Ising model are
\begin{equation}\label{w}
W(\sigma,\sigma^j)= \psi(\sigma,j)\,\exp
-\frac{\beta}{2}[U(\sigma^j) - U(\sigma)]
\end{equation}
and all  produce the same equilibrium \eqref{eq} independent of
the prefactor $\psi(\sigma,j)=\psi(\sigma^j,j)$ as long as it
indeed does not depend on $\sigma(j)$. Moreover there is then a
time-reversal symmetry for the stationary process: denoting by
$P_\rho$
 the stationary  Markov (equilibrium) process with stationary law $\rho$, we have
equilibrium correlations
\begin{eqnarray}\label{cor}
\langle f(\sigma_s)\,g(\sigma_t)\rangle_{\textrm{eq}}  = \langle
f(\sigma_t)\,g(\sigma_s)\rangle_{\textrm{eq}}
\end{eqnarray}
which are functions of $|t-s|$. The brackets $\langle \cdot
\rangle_{\textrm{eq}}$ denote an average over the equilibrium
ensemble over all possible realizations of the stochastic process
determined by \eqref{eq}--\eqref{w}.

\subsection{Perturbation and response}
 Suppose now that we start in equilibrium $\rho$ at time $t=0$ but thereafter
 we slightly modify the dynamics
  in a time-dependent
  way. For times $s \in [0,t]$
  we switch on a
  magnetic
  field of small amplitude  $h_s$.   That is the
external stimulus by which we change the energy function $U$ into $U - h_s\, V$ for $V(\sigma) =\sum_{i\in \Lambda}
\sigma_i$.  How will the equilibrium system respond at time $t > s$, and does the choice of $\psi$ in the rates
 (\ref{w})
make a difference?\\
We look at the linear
response
\[
\langle Q(t)\rangle^h_\rho = \langle Q(t) \rangle_{\textrm{eq}} +
\int_0^t\id s\, h_s R_{QV}^{\textrm{eq}}(t,s) + o(h)\]
 Here, $Q(t) = Q(\sigma_t)$ is a
function of the random spin configuration evaluated at time $t>0$.
The left-hand side averages over the  perturbed dynamics,
depending on the $h_s$, and over the initial equilibrium $\rho$;
the right-hand side averages over the unperturbed dynamics always
starting in $\rho$: $\langle Q(t)\rangle_{\textrm{eq}} =
\sum_\sigma \rho(\sigma)\, Q(\sigma)$ as the equilibrium is
time-invariant. The linear correction contains the response
function or generalized susceptibility $R_{QV}^{\textrm{eq}}(t,s)$
which is our object of study. Formally and leaving away further
decorations,
\[
R_{QV}(t,s) = \left.\frac{\delta}{\delta h_s}\right|_{h=0}\langle
Q(t)\rangle^h
\]
An interesting case looks at the response in the
magnetization itself, taking $Q = V = \sum \sigma(i)$ and then
\begin{equation}\label{vier}
R_{QV}^{\mbox{eq}}(t,s) = \beta\,\sum_{i,j\in \Lambda}\,
\frac{\partial}{\partial s}\langle
\sigma_s(i)\,\sigma_t(j)\rangle_{\textrm{eq}}, \quad 0 < s < t
\end{equation}
is expressible as a space-time--correlation function in the equilibrium
process.  That formula is valid for all times $0< s< t$ \textit{and}
for all choices of rates that satisfy detailed balance.
It is an example of the fluctuation-dissipation theorem for
finite-time perturbations.  The more general equilibrium formula
of which \eqref{vier} is a special case reads
\begin{equation}\label{eqform}
R_{QV}^{\mbox{eq}}(t,s) = \beta\,\frac{\partial}{\partial s}\left<
V(s)Q(t)
 \right>_{\textrm{eq}},\quad 0 < s <t
 \end{equation}
which is again true for any choice \eqref{w} of the rates that
satisfies detailed balance. A proof of this is easy by applying
first-order time-dependent perturbation theory
and by inserting the equilibrium condition \eqref{eq}.\\

If we integrate \eqref{eqform} over $s\in [0,t]$ with constant $h_s=h$, then
\begin{equation}\label{poteq}
\langle Q(t)\rangle^h  - \langle Q(t) \rangle_{\textrm{eq}} =
 h\beta[\langle V(t)Q(t)\rangle_{\textrm{eq}} -
\langle V(0)Q(t)\rangle_{\textrm{eq}}]
\end{equation}
Taking $t\uparrow +\infty$ we recognize the usual change in the equilibrium Boltzmann-factor
to first order in $h$ when changing the potential $U\rightarrow U-hV$.\\

The conclusion in equilibrium: adding a potential to a system with
an equilibrium dynamics is unambiguous, at least when looking at
linear response and there is a simple and explicit linear response
formula in terms of an equilibrium correlation in which we
recognize the Boltzmann factor.

The rest of this paper addresses the question what happens if the
unperturbed dynamics is out-of-equilibrium. The answer is again an
explicit formula (see \eqref{prof1}--\eqref{stationary} below) but
the choice of how to add a potential now does have some influence
on the response formula. Moreover, there is an interesting
interpretation of the resulting correlations in terms of dynamical
fluctuations, which extends equilibrium considerations --- see
Proposition \ref{theorem3}.

\section{Going nonequilibrium }
The extension of the previous problem to a nonequilibrium set-up
has been considered in many papers.  We take here the approach of
\cite{prl,bmw}.\\
We consider a Markov stochastic dynamics for a finite
system. Denote the state space by $K$. We have transition rates
$W(x,y), x,y\in K$. We do no longer assume that there is a
potential, i.e. a function $U(x), x\in K$ for which $W(x,y) \exp -
U(x) = W(y,x) \exp - U(y)$.  In particular, for a stationary
distribution $\rho(x), x\in K$, while
\[
\sum_{y\in K} [\rho(x)\,W(x,y) - \rho(y)\,W(y,x)] = 0, \quad x\in
K \] still, there are nonzero currents of the form
$\rho(x)\,W(x,y) - \rho(y)\,W(y,x)\neq 0$ for some pairs $x\neq
y\in K$. The stationary process (Markov dynamics in $\rho$) is
then no longer time-reversible.  We have in mind systems of
stochastically interacting
particles which are driven away from equilibrium; the state $x$ is then the total configuration
of particles and the transitions are local.  An example follows in Section \ref{ex1}.\\
Secondly, we also do not need to assume that we start at time
$t=0$ from a stationary distribution.  Rather, we have an
arbitrary probability distribution $\mu(x), x\in K,$ from which
the initial data are
drawn and then for $t>0$ we apply the perturbed dynamics.\\

The question is first how to perturb the transition rates
$W(x,y)\to W_t(x,y)$, by adding an extra potential $-h_tV$ to the
system. For convenience we assume that the perturbation $h_s, s >
0,$ is twice differentiable.  Our (physical) assumption here is
that the perturbed rates at time $t>0$ should satisfy
\begin{equation}\label{demand}
 \frac{W_t(x,y)}{W_t(y,x)} = \frac{W(x,y)}{W(y,x)}e^{\beta h_t\,[V(y)-V(x)]}
\end{equation}
The inverse temperature $\beta$ signals that the perturbation concerns an
additional energy exchange with a reservoir at temperature
$\beta^{-1}$.  The assumption \eqref{demand} is conform the condition of local detailed
balance as often applied in particle systems.  That is why we speak of an energy impulse. \\

Condition \eqref{demand} leaves many possible choices for the
perturbed transition rates. A quite general choice is
\begin{equation}\label{general}
 W_t(x,y) = W(x,y)\,e^{h_t[b V(y)-a V(x)]}
\end{equation}
where the $a,b \in
\bbR$ are independent of the potential $V$; it was considered in
\cite{diez}. To satisfy (\ref{demand}) we need that $a+b = \beta$
but $a$ or $b$ can still vary.  A first choice is
\begin{equation}\label{case1}
W_t^{(1)}(x,y) = W(x,y)\,e^{\frac{\beta\, h_t}{2} [V(y) - V(x)]}
\end{equation}
in which case $a= b = \beta/2$; that is sometimes called the
force-model and was explicitly treated in \cite{bmw}. A second
case is
\begin{equation}\label{case2}
W_t^{(2)}(x,y) = W(x,y)\,e^{-\beta\, h_t V(x)}
\end{equation}
whence  $b=0, a=\beta$, or the opposite $b=\beta, a=0$.\\
It is instructive to understand the difference between these cases: we can rewrite
\eqref{general} as
\begin{equation}\label{1general}
W_t(x,y) = W(x,y)\,e^{h_t\,\frac{b-a}{2}[V(y) + V(x)]}\,e^{\frac{h_t\beta}{2}[V(y)- V(x)]}
\end{equation}
and we see that making $a\neq b$ gives an extra $x\leftrightarrow
y$ symmetric but time-dependent prefactor
$\psi_t(x,y)=\psi_t(y,x)=\exp h_t\,\frac{b-a}{2}(V(y) + V(x))$
with respect to the force-model of \eqref{case1}.
 Visualizing the situation in terms of a one-dimensional
 potential landscape we imagine the states $x$  located at the local minima of a potential $U$ and separated from each
 other via energy barriers.
 The Arrhenius formula then predicts a rate $W(x,y) \propto \exp -\beta[D(x,y) - U(x)]$
 where $D(x,y)=D(y,x)$ is the barrier height between states $x$ and $y$.
  Naturally, adding a time-dependent potential
  landscape can affect both the symmetric prefactor $D(x,y)$ and
   the local minima $U(x)$ themselves which gives a possible
   interpretation of the two constants $a$ and $b$.  For example, choosing \eqref{case2} only changes the depth of the local minima (binding energies $U(x) \rightarrow U(x) - h_t V(x)$) and not the barrier heights.
The above picture works best under equilibrium conditions, but one now imagines
that the nonequilibrium driving adds further asymmetries.\\

The linear response question remains unchanged:  at time $t>0$ the
expected value of an observable $Q$ will probably deviate from the
expectation under the unperturbed dynamics. Linear response theory
out-of-equilibrium is interested in estimating and interpreting
the deviations
\[
\langle Q(t)\rangle^h_\mu - \langle Q(t)\rangle_\mu
\]
to first order in $h$.  We have abbreviated $Q(t) = Q(x_t)$ for
the observable at  time $t$.  In other words, we want to compute
$R_{QV}^\mu(t,s) = R(t,s), 0 < s < t$, in
\[
\langle Q(t)\rangle^h_\mu = \langle Q(t) \rangle_{\mu} +
\int_0^t\id s\,
h_s\, R(t,s) + o(h)\]\\

\section{Response formula}

The present section computes the response function $R(t,s)$ for
the general perturbation of the form (\ref{general}).  In the
formula appears the backward generator $L$ of the jump process; in
terms of the transition rates,
\[
Lf(x) = \left.\frac{\id}{\id s}\right|_{s=0}\langle
f(x_s)\rangle_{x_0=x}\; = \;\sum_y W(x,y)[f(y)-f(x)]
\]

\begin{proposition}\label{theorem1}
 For a perturbation of the form (\ref{general}), the response function is equal to
\begin{eqnarray}\label{prof1}
 R(t,s) =&& b\frac{\partial }{\partial s}\left<V(x_s)Q(x_t)\right>_{\mu} -
  a\frac{\partial }{\partial t}\left<V(x_s)Q(x_t)\right>_{\mu}\nonumber\\
+ && b\left[ \left<V(x_s)LQ(x_t)\right>_{\mu} -
\left<LV(x_s)Q(x_t)\right>_{\mu} \right]
\end{eqnarray}
\end{proposition}

The proof  of this result is essentially a linear order
perturbation of the Girsanov-formula for the density of the
perturbed versus the original path-space measures.

\begin{proof}[Proof of Proposition \ref{theorem1}]
 To see where we must
go, we first rewrite the right-hand side of \eqref{prof1}. In
particular, the third term involving $LQ$ can directly be combined
with the second term, time-derivative in $t$, adding up to $(b-a)$
multiplied with
\[
\frac{\partial}{\partial t} \langle V(x_s) Q(x_t)\rangle_\mu = -
\frac{\partial}{\partial s}\langle V(x_s)\,Q(x_t)\rangle_\mu +
\sum_x \dot{\mu}_s(x)\,V(x) e^{(t-s)L}Q(x)
\]
where $\dot{\mu}_s(x) = \sum_y W(y,x)\mu_s(y) - \sum_y W(x,y)
\mu_s(x)$ solves the master equation starting from $\mu_0=\mu$. As
a consequence, we really must prove that
\begin{eqnarray}\label{tp}
R(t,s) &=& a \,\frac{\partial}{\partial s}\langle V(x_s)
Q(x_t)\rangle_\mu \nonumber\\
- b\langle LV(x_s) Q(x_t)\rangle_\mu &-& (b-a)\, \sum_y \langle
W(x_s,y) V(x_s)
Q(x_t)\rangle_\mu \\
&+& (b-a)\,\sum_{x,y} \mu_s(y) W(y,x) V(x) e^{(t-s)L}Q
(x)\nonumber
\end{eqnarray}
Let a path be denoted by $\omega = (x_s)_s, s\in [0,t]$, for
$x_s\in K$. Paths are piecewise constant and chosen with left
limits and right continuous at every jump time.  For the perturbed
process
\begin{equation}\label{expectation}
 \langle Q(t)\rangle^h_\mu = \int\,\id P_{\mu}(\omega)\,  \frac{\id
P^h_\mu}{\id P_\mu}(\omega) \,Q(x_t)
\end{equation}
where we have inserted the  density between the path-measures
$P_{\mu}^h(\omega)$ for the perturbed and the unperturbed ($h=0$)
Markov dynamics starting from law $\mu$ at time zero. Explicitly (see e.g. Appendix 2 in \cite{KL}),
the Girsanov formula gives
\begin{eqnarray}\label{pro} \log \frac{\id P^h_\mu}{\id
P_\mu}(\omega)
 &=&  \sum_{s\leq t}\,h_s\,\big[b V(x_s)- a V(x_{s^-})\big]\nonumber\\
 && - \sum_y\int_0^t\id s \,W(x_s,y)\big[e^{h_s[bV(y)-aV(x_s)]}-1\big]
\end{eqnarray}
where the first sum is over all the jump times $s \in [0,t]$.
Up to linear order in $h$, and with some reordering of the terms, this becomes
\begin{eqnarray}\label{pro1}
\log \frac{\id P^h_\mu}{\id P_\mu}(\omega)
&=&  (b-a)\sum_{s\leq t} h_s V(x_{s}) + a \sum_{s\leq t} h_s [V(x_{s}) - V(x_{s^-})] \nonumber\\
&& - b\int_0^t \id s \,h_s\, LV(x_s) - (b-a)\int_0^t \id s\, h_s \sum_{y} W(x_s,y) V(x_s)\nonumber\\
\end{eqnarray}
Higher order in $h$ can easily be controlled.
 Its  second term on the right
still allows a partial summation into
\begin{eqnarray}
\sum_{s\leq t}\,h_{s}\,\big[V(x_{s})- V(x_{s^-})\big] &=& h_t
V(x_t) - h_0 V(x_0) - \sum_{s\leq t} V(x_{s^-}) \, \big[ h_{s} -
h_{s^{-}}\big ]\nonumber\\
 = h_t V(x_t) - h_0 V(x_0) &-& \int_0^t \id s\,\frac{\id}{\id s}
h_s \; V(x_s)
\end{eqnarray}
The expression \eqref{pro1} must now be multiplied with $Q(x_t)$
and averaged over the original Markov process starting from $\mu$,
after which we note that
\begin{equation}\label{wo} \langle
\{\big[h_t V(x_t) - h_0 V(x_0)\big] - \int_0^t \id
s\,\frac{\id}{\id s} h_s \; V(x_s) \}\,Q(x_t)\rangle_\mu =
\int_0^t\id s\, h_s \frac{\partial}{\partial s}\langle V_s
Q_t\rangle_\mu \end{equation}
 reproduces the first term in
\eqref{tp}. The last two terms in \eqref{pro1} are also easily
identified giving rise to the two middle terms in \eqref{tp}. That
leaves us with the very first term in \eqref{pro1} for which must
hold that
\[
\langle \sum_{s\leq t} h_s\,V(x_s)\,Q(x_t)\rangle_\mu = \sum_{x,y}
\int_0^t \id s \,h_s\, \mu_s(y) W(y,x)\,V(x) e^{(t-s)L}Q(x)
\]
That is indeed true as can be seen by writing  the sum over all
jump times in terms of the random measure $\id k_s(y,x)$ on paths
$\omega$, which gives 1 when there is a jump $y\rightarrow x$ at
time $s$, and is zero otherwise:
\[
\langle \sum_{s\leq t} h_s\,V(x_s)\,Q(x_t)\rangle_\mu = \sum_{x,y}
V(x) \int_0^t h_s\,\langle \id k_s(y,x)\,  Q(x_t)\rangle_\mu
\]
By the Markov property $e^{(t-s)L}Q(x) = \langle Q(x_t)|x_s=x,
x_{s^-}=y\rangle_\mu$ and
\[
\langle \id k_s(y,x)\, Q(x_t)\rangle_\mu = \mu_s(y)\,W(y,x)
e^{(t-s)L}Q(x)\,\id s
\]
so that the conclusion \eqref{prof1} is reached.
\end{proof}

\section{Example}\label{ex1}
 We come back to the example \eqref{w} of a purely dissipative spin-flip dynamics. We now add a
mixing dynamics.
 More specifically, we not only have transitions $\sigma\rightarrow \sigma^j$ with
 corresponding rates $W(\sigma,\sigma^j)$, but now we also allow transitions
 $\sigma \rightarrow \sigma^{ij}$ where the spins at neighboring vertices $i \sim j\in \Lambda$ get
 exchanged: $\sigma^{ij}(k) = \sigma(k),$ if $i\neq k\neq j$ while
  $\sigma^{ij}(i)=\sigma(j), \sigma^{ij}(j)=\sigma(i)$.  The
  rate for these exchanges is $\lambda>0$.
The result is a reaction-diffusion process on $K =
\{+1,-1\}^\Lambda$ with generator $L$ acting on functions $f: K\rightarrow
\bbR$,
\[
Lf(\sigma) = \sum_{j\in \Lambda}W(\sigma,\sigma^j)[f(\sigma^j) -
f(\sigma)] +\lambda \sum_{i\sim j}[f(\sigma^{ij}) - f(\sigma)]
\]
That unperturbed dynamics does not satisfy the condition of
detailed balance when $\beta\neq 0$ for a nontrivial energy
function $U(\sigma)$ in \eqref{w}.  There is a stationary distribution $\rho$
of which very little is known; in particular it can depend on the $\psi$ in \eqref{w}.\\
 We still consider the magnetization
$V(\sigma)=Q(\sigma) = \sum_i \sigma_i$ for organizing and
evaluating the perturbation of amplitude $h_t, t>0$.  Note that
$V(\sigma^{ij}) = V(\sigma)$ and the transition $\sigma
\rightarrow \sigma^{ij}$ leaves the total magnetization unchanged.
Hence $LV(\sigma) = -2\sum_i \sigma_i W(\sigma,\sigma^i)$ is still the
dissipation of magnetization due to the spin flip reaction. Let us
abbreviate $J_i(\sigma) = -2\sigma(i)\,W(\sigma,\sigma^i)$ for the systematic
rate of change in the local magnetization. We get the linear
response around steady nonequilibrium from \eqref{prof1}:
\[
\left.\frac{\partial}{\partial h_s(i)}\right|_{h=0}\langle\sigma_j(t)\rangle_\rho^h =
 a\, \frac{\partial}{\partial s}\langle
\sigma_s(i)\,\sigma_t(j)\rangle_\rho - b\, \langle J_i(\sigma_s)\,\sigma_t(j)\rangle_\rho
\]
We see that the equilibrium expression \eqref{vier} gets modified
by the correlation between $\sigma_t(j)$ and the flux $J_i(\sigma_s)$.
For a constant perturbation $h_s=h, s\in [0,t]$, we can integrate
over $s\in [0,t]$ to get the leading order of the response:
\begin{eqnarray}\label{redi} \frac 1{h}\sum_i
\langle\sigma_t(i)\rangle^h_\rho - \langle\sigma_0(i)\rangle_\rho
&=& a\,\sum_{i,j\in \Lambda}\,\langle [\sigma_t(i)
-\sigma_0(i)]\,\sigma_t(j)\rangle_\rho\\
  &-&
b \,\sum_{i,j\in \Lambda}\int_0^t\id s\langle J_i(\sigma_0)\,\sigma_s(j)\rangle_\rho\nonumber
\end{eqnarray}
Note that the rate $\lambda$ is hiding in the correlation
functions but the form \eqref{redi} is unchanged no matter what is
$\lambda$.\\
  The example is a more microscopic version of a reaction-diffusion model
  but it
can also be considered as a toy model for a granular lattice gas
undergoing inelastic collisions. The spins refer then to the
presence or absence of energy packets which diffuse but can also
be created or get lost.  The latter specifies the temperature of
the environment and the flux $J_i$ in the above would be the
systematic rate of local energy change.

\section{More symmetries}

In a number of cases the response formula \eqref{prof1}
simplifies.\\
 There is first the case where the initial distribution
is the stationary measure $\rho$. Then, i.e., when $\mu=\rho$ is
the stationary distribution, correlation functions like $\langle
V(x_s)Q(x_t)\rangle_{\rho}$ are functions of $t-s$, so that the
response function becomes
\begin{equation}\label{stationary}
 R_{QV}(t,s) =
 a\frac{\partial}{\partial s}\langle V(x_s)Q(x_t)\rangle_{\rho}
 - b\langle LV(x_s)Q(x_t)\rangle_{\rho}
\end{equation} In equilibrium, i.e., under time-reversal
symmetry, the two terms in the right-hand side of
\eqref{stationary}
coincide and we recover \eqref{eqform} whenever $a+b=\beta$ (and independent of $\psi$ in \eqref{w}).\\

For the case (\ref{case1}),
 this means that $b=a=\frac{\beta}{2}$, the response formula becomes
\[ R_{QV}(t,s) = \frac{\beta}{2}
\frac{\partial}{\partial s}\langle V(x_s)Q(x_t)\rangle_{\mu}-\frac{\beta}{2}\langle LV(x_s)Q(x_t)\rangle_{\mu}\]
which is in exact agreement with \cite{prl}.\\

A special case arises when $b=0$ and $a=\beta$ in (\ref{general}),
because then the response is of the same form as in equilibrium:
\[ R(t,s) = -\beta\frac{\partial}{\partial t}\langle V(x_s)Q(x_t)\rangle_{\mu} \]
This is indeed a special kind of perturbation, as can also be seen
from the following consideration. Take $h$ to be constant; the law
$\rho^h$ defined by $\rho^h(x) \propto \rho(x)e^{\beta h V(x)}$ is
stationary for the new dynamics (to all orders in $h$).  In other
words, here the resulting behavior under this perturbation is like
in equilibrium, even though the unperturbed dynamics can be far
from equilibrium.\\

That last remark brings us to  considering the limit $t\uparrow
\infty$ of \eqref{prof1} in which the response formula should show
stationary response. Imagine thus that we apply a new
time-independent dynamics with rates
\begin{equation}\label{pers}
W^V(x,y) = W(x,y)\,\,e^{h[b V(y)-a V(x)]},\; \textrm{ small
constant } h
\end{equation}

We assume that both the original ($h=0$) and the perturbed
dynamics show exponential ergodicity in converging to $\rho$,
respectively $\rho^V$.  Similarly we can replace in the above the
function $V$ by another function $M$ on $K$ and construct
$\rho^M$. Both $\rho^V$ and $\rho^M$ depend on $h$ and we
investigate their change with respect to the original $\rho$ to
first order in $h$. The original backward generator is still $L$.
The following proposition looks at a special observable, and we
write
\begin{equation}\label{srf}
\chi_{MV}^{ab}=\left.\frac{\delta}{\delta h}\right|_{h=0} \sum_x
\rho^V(x) LM(x);\quad \chi_{VM}^{ab}=\left.\frac{\delta}{\delta
h}\right|_{h=0} \sum_x \rho^M(x) LV(x)
\end{equation}
The dependence on the constants $a,b,h$ is not made explicit but
sits in the perturbed dynamics, as in \eqref{pers} for perturbing
potential $V$ and similarly for perturbation $M$.

\begin{proposition}\label{theorem2}
The stationary response functions \eqref{srf} equal
\begin{equation}\label{statio} \chi_{MV}^{ab}= \chi_{VM}^{ba} =
b\langle M\,LV\rangle_\rho + a\langle V\,LM\rangle_{\rho}
 \end{equation}
\end{proposition}

Observe the symmetry  when interchanging $M$ and $V$ together with
the exchange of $a$ and $b$. If the perturbation is of the form
(\ref{case1}), then only interchanging $M$ and $V$ is enough. This
symmetry appears useful because it reduces the amount of response
functions to be measured. Moreover, some experimentally difficult
responses can be made more accessible by interchanging the role of
observable and perturbation.  While its proof is trivial, we are
not aware that this symmetry \eqref{statio} has been observed
before. On the level of generators it simply amounts to the direct
observation that
\begin{equation}\label{lid}
(L^V_{ab}-L)M = (L^M_{ba}-L)V + h(b-a)\,L(MV) + O(h^2)
\end{equation}
where for example $L^V_{ab}f(x) = \sum_y W(x,y) \exp[ h(bV(y) -
aV(x)) ][f(y) - f(x)]$.  The symmetry in \eqref{statio} then
easily follows from averaging the identity \eqref{lid}. We add
however a different proof that connects with the response formula
\eqref{prof1}.

\begin{proof}[Proof of Proposition \ref{theorem2}]
 For observables $Q$ which are of the form $Q = LM$,
the linear response is given by \eqref{prof1},
\begin{equation}\label{stationaryL}
 R_{LM,V}^\rho(t,s) = \frac{\partial}{\partial s}
 [a\langle V(x_s)LM(x_t)\rangle_{\rho} + b\langle LV(x_s)M(x_t)\rangle_{\rho}]
\end{equation}
When we consider a constant perturbation $h_s=h$, we get the
integrated form of the response function:
\begin{eqnarray*} R(t) &=& \int_0^t\id s R_{LM,V}^\rho(t,s) \\
&=& a\langle [V(x_t)-V(x_0)]LM(x_t)\rangle_{\rho} +
 b\langle [LV(x_t)-LV(x_0)]M(x_t)\rangle_{\rho}
  \end{eqnarray*}
It suffices to take $t\to\infty$ to see the appearance of
\eqref{statio}.  The exchange of the limits $h\rightarrow 0$ and
$t\uparrow +\infty$ is trivial in the case considered, so that for
constant $h_t=h$
\[ \lim_t R(t) = \chi_{MV}^{ab}
\]
\end{proof}

\section{Response and dynamical fluctuations}
We turn to the interpretation of the response functions in terms
of fluctuation theory.  The standard interpretation of the
equilibrium response \eqref{eqform} is in terms of energy
dissipation, and that is why \eqref{eqform} is called  the
fluctuation-dissipation theorem even though it really deals with
response.  That terminology and corresponding interpretation
remains true and useful for the first term in \eqref{stationary}
at least when considering the flux in excess to what already was
present (since we now deal with nonequilibrium). That was
explained in \cite{bmw}, section 5, and we also see it in the
identity \eqref{wo} which can be interpreted as a conservation of
energy. From a probabilistic point of view it is more interesting
to concentrate on what is new with
respect to equilibrium, the second term in \eqref{stationary}.\\

The way of responding and the way of fluctuating are like each
other's time-reversals.  For inspiration, we turn again to the
equilibrium (hence, time-reversible) case, where the response to a
perturbation typically goes along the same path as that of a
spontaneous fluctuation; that is sometimes called Onsager's
regression hypothesis and in our context it could be summarized as
$L=L^*$ where $L^*$ is the adjoint in the $\rho$-scalar product.
In nonequilibrium the regression of a fluctuation
is also the time-reversal of its appearance, but now the time-reversal is not trivial.
In particular the second term in \eqref{stationary} is
\[
-b\,\sum_x \rho(x) \,V(x)\,L^*e^{(t-s)L}Q(x)
\]
which now cannot be written as a time-derivative as in the first
term of \eqref{stationary}.  Yet, it is related to a fluctuation, as we explain now.\\
Suppose a (constant) perturbation $V$ is added to the system.  The
system responds but in the long time, the perturbation also
installs a new stationary law. Let us denote this new stationary
law by $\mu$. On the other hand one can compute the probability
that in the unperturbed dynamics $\mu$ occurs as a fluctuation.
This takes us to the dynamical fluctuation theory for Markov
processes, started by \cite{DV}, see also e.g. in \cite{DZ}.
Without going to the full details it suffices here to recall that
for an ergodic Markov process with backward generator $L$ there is
a fluctuation functional $I(\mu)$ on the probability laws $\mu$ on
$K$, of the form
\begin{equation}\label{imu}
I(\mu) = - \inf_{g>0}  \sum_x \mu(x) \frac{Lg}{g}(x)
\end{equation}
for which in the sense of the theory of large deviations
\[
\mbox{Prob}_\rho[p_\tau\simeq \mu] \simeq e^{-\tau I(\mu)},\quad
\tau \uparrow +\infty
\]
for the empirical distribution
\[
p_\tau(x) = \frac 1{\tau}\,\int_0^\tau \delta_{x_t,x}\,\id t,\quad
\mbox{with}\; \delta_{a,b} = 0 \mbox{ if } a\neq b \;\mbox{and}\;
\delta_{a,b}= 1 \mbox{ if } a = b
\]
of occupation times over the time-interval $[0,\tau]$. We refer to
\cite{mnw} for a dynamical fluctuation theory in the context of
the present paper. In a sense, $\exp -\tau I(\mu)$ gives the
plausibility of the long-term ($=\tau$) appearance of (= dynamical
fluctuation to) the statistics $\mu$. Taking $g = \exp(b hM/2)$ in
\eqref{imu}, we see
\begin{equation}\label{dv}
I(\mu) = - \inf_{M} \left\{ \sum_x \mu(x) \left[\sum_y W(x,y)
e^{\frac{bh}{2}[M(y)-M(x)]} - \sum_y W(x,y)\right] \right\}
 \end{equation}
That $M$ can now be interpreted as a potential. The infimum in
\eqref{dv} gets reached at $M=V$, the potential for which $\mu$ is
the stationary law. Already here we see a complementarity between
response and fluctuations: a perturbation $V$ gives a new
stationary law $\mu$, and to find the probability of a fluctuation
$\mu$ in the original dynamics, one has to find exactly this $V$.
But there is also a quantitative relation, in particular as
realized in the second term in \eqref{stationary}, as we prove in
the next proposition.

\begin{proposition}\label{theorem3}
The dynamical fluctuation functional $I(\mu)$ satisfies
\begin{equation}\label{res3}
I(\mu) =  -\frac{bh}{4}\sum_x\mu(x)\,LV(x) + o(h^2)
\end{equation}
\end{proposition}

\begin{proof}[Proof of Proposition \ref{theorem3}]
For our case we can exchange the limit by the infimum and the
small $h$ limit, see also \cite{mn}.
 We can then compute \eqref{imu} to first order
  in $h$ by taking $M=bhV/2$ in \eqref{dv}, and expanding
   \begin{eqnarray}\label{axp}
&&\sum_{x,y} \mu(x)\, W(x,y)\,[1 - e^{b\, h[V(y) -
  V(x)]/2}] =
  -\frac{bh}{2}\sum_{x,y}\mu(x)\,W(x,y)[V(y)-V(x)]\nonumber\\
&&-\frac{b^2h^2}{8}\sum_{x,y}\rho(x)\,W(x,y)\,(V(y)-V(x))^2 +
o(h^2)
\end{eqnarray}
where we have used already that $\langle LV\rangle_\rho =0$ by
stationarity.   For the second term we can replace the $\rho(x)$
by $\mu(x)$ because we are already at second order in $h$ and
write
\begin{eqnarray*}
&&-\frac{b^2h^2}{8} \sum_{x,y}\mu(x)\,W(x,y)\,(V(y)-V(x))^2 + o(h^2)=\\
&&\frac{bh}{4}\sum_{x,y}\mu(x)\,W(x,y)[V(y)-V(x)]\,[1-e^{b\,
h[V(y)
- V(x)]/2} ] \nonumber\\
&=& \frac{bh}{4}\sum_{x,y}\mu(x)\,W(x,y)[V(y)-V(x)]
\end{eqnarray*}
because $\mu$ is invariant under the perturbed dynamics.
Collecting all terms we get the result \eqref{res3}.

\end{proof}

Of course the $LV(x)$ in the response formul{\ae}
\eqref{prof1}--\eqref{stationary}--\eqref{statio} has the usual
meaning of being the expected rate of change in $V$ while at $x$.
Proposition \ref{theorem3}  adds the interpretation that it can
also be seen as the change in escape rate from $x$ when adding a
potential $V$.  From \eqref{res3} the second term in
\eqref{stationary} gives a correlation with a generalized escape
rate and thus relates with the dynamical fluctuations of the
occupation times, \cite{mnw,mn}.\\

\section{Conclusion}

We have generalized the results of \cite{prl,bmw} to the
perturbation first considered by \cite{diez}, Proposition
\ref{theorem1}. We have also added a stationary response relation
and noted a new symmetry, Proposition \ref{theorem2}. The
fluctuation interpretation of \cite{bmw} remains intact,
Proposition \ref{theorem3}.\\

\vspace{2cm}

\noindent {\bf Acknowledgments:} We are grateful to Marco Baiesi
and Karel Neto\v{c}n\'{y} for useful discussions. B.W. is aspirant
(research assistant) at the Flemish science foundation FWO.
Support from the Belgian Interuniversity Attraction Pole P6/02 is
also acknowledged.

\end{document}